\documentclass[10pt,prl,aps,twocolumn,preprintnumbers,showpacs, nofootinbib,superscriptaddress,notitlepage]{revtex4-2}

\usepackage[T1]{fontenc}
\usepackage{graphicx}
\usepackage{amsmath}
\usepackage{amsmath}
\usepackage{xspace}
\usepackage{subfig}
\usepackage{color}
\usepackage[utf8]{inputenc}
\usepackage{commath}
\usepackage{cancel}
\usepackage{slashed}
\usepackage{fancyvrb}
\usepackage{multirow}
\usepackage{hyperref}
\usepackage{tikz}
\usepackage{fancyhdr}
\usepackage[normalem]{ulem}

\begin{document}

\title{Testing Leptogenesis from Observable Gravitational Waves}

\author{Wei Liu}
\email{wei.liu@njust.edu.cn}
\affiliation{ Department of Applied Physics and MIIT Key Laboratory of Semiconductor Microstructure and Quantum Sensing,
Nanjing University of Science and Technology, Nanjing 210094, China}

\author{Yongcheng Wu}
\email{contact author: ycwu@njnu.edu.cn}
\affiliation{Department of Physics, Institute of Theoretical Physics and Institute of Physics Frontiers and Interdisciplinary Sciences, Nanjing Normal University, Nanjing, 210023, China}
\affiliation{Nanjing Key Laboratory of Particle Physics and Astrophysics, Nanjing, 210023, China}

\begin{abstract}
Leptogenesis provides an elegant mechanism to explain the observed baryon asymmetry of the Universe (BAU), yet its experimental verification remains challenging due to requirements of either extremely heavy right-handed neutrinos or precisely fine-tuned mass splittings.
We present a scenario where an additional scalar field, coupled to both the Higgs and right-handed neutrinos (RHNs), enhances the CP asymmetry through loop-level contributions, enabling successful leptogenesis at the electroweak scale. This same scalar sector triggers a strong first-order electroweak phase transition (EWPT), producing gravitational waves (GWs) within the reach of next-generation detectors. We demonstrate a robust correlation between the BAU and GW signal strength, providing a falsifiable prediction of leptogenesis that future GW searches can confirm or rule out.
\end{abstract}
\maketitle
\preprint{$\begin{gathered}\includegraphics[width=0.04\textwidth]{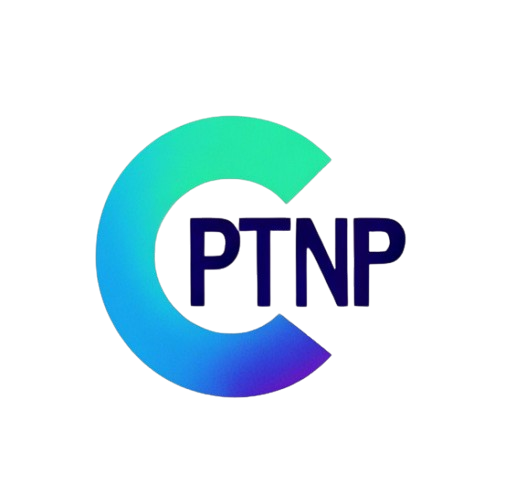}\end{gathered}$\,CPTNP-2025-001}

\section{Introduction}
\label{sec:intro}

The observed baryon asymmetry of the Universe (BAU) remains one of the most profound unsolved puzzles in modern particle physics and cosmology. Among the proposed explanations, leptogenesis driven by $CP$-violating decays of right-handed neutrinos (RHNs) stands out as a particularly compelling mechanism~\cite{Fukugita:1986hr,Luty:1992un,Davidson:2008bu}. This scenario is naturally embedded within the framework of seesaw mechanisms that simultaneously address the origin of small neutrino masses~\cite{Minkowski:1977sc}\footnote{For recent developments involving RHNs and seesaw mechanisms, see Refs.\cite{Deppisch:2018eth, Amrith:2018yfb, Deppisch:2019kvs, Chiang:2019ajm, Liu:2021akf, Liu:2022kid, Liu:2022ugx, Zhang:2023nxy, Liu:2023nxi, Barducci:2023hzo, Deppisch:2023sga, Liu:2023klu, Liu:2024fey, Wang:2024mrc, Wang:2024prt, Liu:2025ldf, Bolton:2025tqw}.}. Despite its elegance, conventional leptogenesis faces several significant challenges. In the hierarchical RHN mass regime, successful baryogenesis typically requires RHN masses $m_N \gtrsim 10^9$ GeV, known as the Davidson-Ibarra bound~\cite{Davidson:2002qv}. While this scale can be lowered to $m_N \gtrsim 10^6$ GeV when flavor effects are included~\cite{Hambye:2003rt,Blanchet:2008pw,Moffat:2018wke}, such high-scale RHNs still introduce large radiative corrections to the Higgs mass, exacerbating the electroweak hierarchy problem~\cite{Vissani:1997ys}. A promising alternative is low-scale leptogenesis with much lighter RHNs, which avoids destabilizing the Higgs sector. Enhanced $CP$ asymmetries can be achieved if at least two RHNs are nearly degenerate in mass, as in resonant~\cite{Pilaftsis:1997jf,Pilaftsis:2003gt} or Akhmedov–Rubakov–Smirnov (ARS)~\cite{Akhmedov:1998qx} leptogenesis, allowing RHN masses near the electroweak scale. Nevertheless, these mechanisms typically require fine-tuned mass splittings between the RHN pairs, challenging both naturalness and model simplicity.

Recent work has shown that a simple extension of the Standard Model (SM) scalar sector by an additional real singlet can enable successful leptogenesis with light RHNs without requiring strong mass degeneracy~\cite{LeDall:2014too, Alanne:2018brf, Borah:2021qmi, Barreiros:2022fpi, Ghosh:2024mpz, Bhattacharya:2024ohh}. Simultaneously, such a singlet can couple significantly to the SM Higgs and modify the electroweak phase transition (EWPT) dynamics~\cite{Liu:2021jyc,Liu:2022nvk,Liu:2024fly,Gould:2024jjt,Aboudonia:2024frg}. If the EWPT becomes strongly first-order, it produces stochastic gravitational waves (GWs) potentially observable at future space-based interferometers such as LISA~\cite{Caprini:2015zlo,LISA:2024hlh}, BBO~\cite{Crowder:2005nr}, TianQin~\cite{TianQin:2015yph,Hu:2017yoc}, Taiji~\cite{Hu:2017mde}, and DECIGO~\cite{Kawamura:2011zz,Kawamura:2006up}. This intriguing connection suggests that the gravitational wave signature from the EWPT could serve as a probe of leptogenesis in scalar-extended scenarios.

In this work, we propose a distinct framework in which the $CP$ asymmetry required for leptogenesis is dominantly generated via scalar-assisted processes, without fine-tuned mass degeneracy.

Importantly, this mechanism is naturally accommodated within the context of seesaw models: once RHNs are introduced to generate neutrino masses, scalar-RHN couplings of the form $SNN$ are theoretically well-motivated (representing additional contributions to Majorana masses), and the same scalar field can naturally induce a first-order phase transition.
This generic interplay—relying on scalar-RHN interactions rather than fine-tuned RHN masses—ensures broad applicability across a wide class of beyond-the-Standard-Model (BSM) theories with extended scalar and neutrino sectors.

To demonstrate this mechanism, we adopt a minimal scalar extension of the SM as a benchmark. In this setup, a strong first-order EWPT driven by large scalar couplings induces sizable thermal corrections to the singlet mass, enabling the decay $N_2 \rightarrow N_1 S$ during the transition. This kinematically opens loop-level $CP$-violating contributions from both vertex and self-energy diagrams, further amplified by the same scalar interactions that strengthen the phase transition. Crucially, the baryon asymmetry $\eta_B$ and the GW signal originate from a common dynamical source, exhibiting an intrinsic, approximately linear correlation.

This direct connection distinguishes our framework from existing approaches involving relativistic bubble wall dynamics~\cite{Baldes:2021vyz, Huang:2022vkf, Borah:2022cdx, Chun:2023ezg}, inflationary GWs~\cite{Datta:2022tab, Chianese:2024nyw, Samanta:2025jec}, cosmic defects~\cite{Dror:2019syi, Barman:2022yos, Chianese:2024gee, Datta:2024tne, Murayama:2025thw}, or early matter-dominated eras~\cite{Blasi:2020wpy, Samanta:2021zzk}, where the BAU and GW signals are largely independent or related with intermediate steps. Our framework thus provides a novel, testable pathway whereby future GW detectors can directly probe the mechanism behind the baryon asymmetry of the Universe, as we discuss further below. Importantly, while we focus on the singlet scalar as a benchmark, our mechanism can be broadly applicable to a wide class of scalar-extended models.


\section{Theoretical Framework and Thermal History}

We consider an extension of the SM with two RHNs, $N_{1,2}$ and a real scalar singlet $\mathcal{S}$. The relevant interactions are:
\begin{align}
\label{eq:LagHE}
-\mathcal{L} & \supset h_{\alpha i}\bar{\ell}N_iH + \frac{1}{2}(M_{ij}+a_{ij}\mathcal{S})N_iN_j + {\rm h.c.},
\end{align}
with $i,j=1,2$ and $\alpha=e,\mu,\tau$. As a benchmark, the scalar sector is minimal extended with the scalar potential given by:
\begin{align}
    V_0(H,\mathcal{S}) &= \mu_H|H|^2 + \lambda_H|H|^4 \nonumber \\
    &\qquad + \frac{1}{2}\mu_S^2 \mathcal{S}^2+\frac{1}{4}\lambda_S\mathcal{S}^4 + \lambda_{SH}\mathcal{S}^2|H|^2.
\end{align}
where $H = (\phi^+, \frac{\omega_h+h+i\phi^0}{\sqrt{2}})^T$, $\mathcal{S}=\omega_s+S$ and $\omega_{h,s}$ are the vacuum expectation values (vevs) for the doublet and singlet respectively. At zero temperature, only the Higgs acquires a vev $\omega_h=v\approx 246\,\rm GeV$, while $\omega_s=0$. There is thus no singlet-Higgs mixing, and collider constraints are avoided if $m_S>m_h$.

\begin{figure}[t!]
\centering
\includegraphics[width=0.45\textwidth]{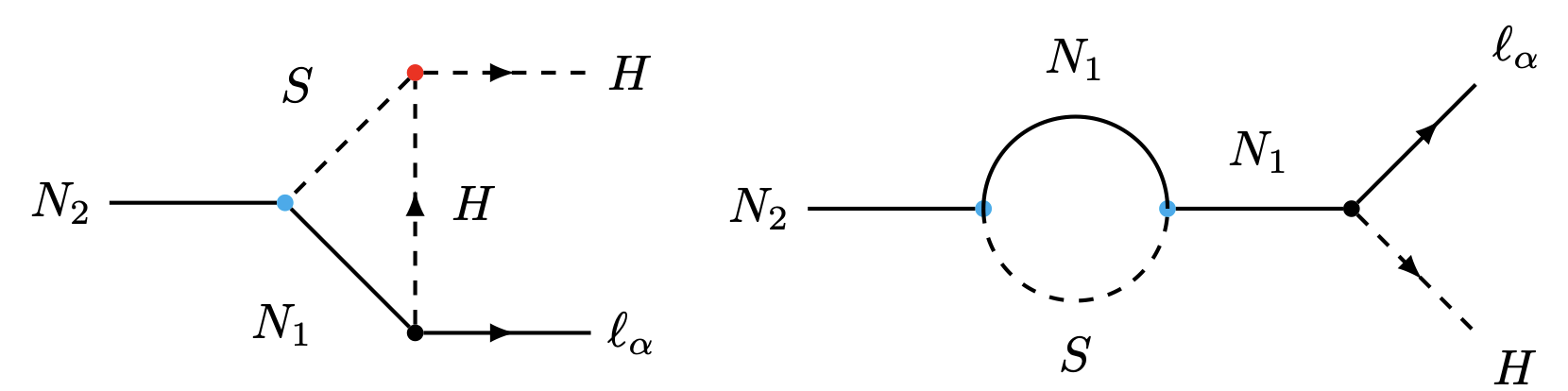}
\caption{\label{fig:N_decay}
The vertex correction~(left) and self-energy diagrams~(right) as the additional sources of $CP$ asymmetry mediated by the scalar $S$ in the loop level, in the decay $N_2 \to \ell_\alpha H$.}
\end{figure}

These interactions generate neutrino masses via the seesaw mechanism and mediate CP-violating decays essential for leptogenesis, as shown in Fig.~\ref{fig:N_decay}. However, the contributions from these diagrams heavily depend on the thermal history of the scalar sector that influences the scalar mass $m_S$ and the trilinear coupling $\mu_{hhs}$ between the Higgs and the singlet.

The thermal evolution is determined by the effective potential at finite temperature. While the analytical expressions listed below use the high-temperature approximation for clarity, all numerical results employ full finite-temperature calculations with higher-order corrections following Refs.~\cite{Lofgren:2021ogg,Hirvonen:2021zej,Brauner:2016fla,Schicho:2021gca,Niemi:2021qvp}. The potential is approximated as
\begin{align}
    V(\omega_h,\omega_s,T) = V_0 + \frac{1}{2}c_hT^2\omega_h^2 + \frac{1}{2}c_sT^2\omega_s^2,
\end{align}
where
\begin{align}
    c_h = \frac{3g^2+g'^2+4y_t^2}{16} + \frac{\lambda_H}{2} + \frac{\lambda_{SH}}{12},\,c_s=\frac{\lambda_S}{4}+\frac{\lambda_{SH}}{3}.
\end{align}
The model allows four vacua: phase-O $(\omega_h,\omega_s)=(0,0)$, phase-S $(0,\omega_s)$, phase-H $(\omega_h,0)$ and phase-SH $(\omega_h,\omega_s)$. A strong first-order EWPT can occur via a two-step transition: $(0,0)\to(0,\omega_s)\to(\omega_h,0)$ when satisfying the following necessary conditions:
\begin{align}
    \frac{c_s}{c_h}<\frac{\mu_S^2}{\mu_H^2}<\frac{\sqrt{\lambda_S}}{\sqrt{\lambda_H}}<\frac{\lambda_{SH}}{\lambda_H}.
\end{align}
The scalar masses and the trilinear coupling $\mu_{hhs}$, will evolve with temperature as
\begin{align}
&\begin{array}{cccc}
       T   & [0,T_{n}] &  \left[T_{n},\sqrt{ -\frac{\mu_{S}^{2}}{c_{s}}}\right] & \left[\sqrt{ -\frac{\mu_{S}^{2}}{c_{s}}},\infty\right] \vspace{1ex} \\
    m_h^2 & -2\tilde{\mu}_{H}^{2} & \frac{\lambda_{S}\tilde{\mu}_{H}^{2}-\lambda_{SH}\tilde{\mu}_{S}^{2}}{\lambda_{S}} & \tilde{\mu}_{H}^{2} \\
    m_S^2 & \frac{\lambda_{H}\tilde{\mu}_{S}^{2}-\lambda_{SH}\tilde{\mu}_{H}^{2}}{\lambda_{H}} & -2\tilde{\mu}_S^2 & \tilde{\mu}_{S}^{2} \\
    \mu_{hhs} & 0 & 2\lambda_{SH}\sqrt{ - \frac{\tilde{\mu}_{S}^{2}}{\lambda_{S}} } & 0
\end{array}
    \label{eq:thermal}
\end{align}
where $\tilde{\mu}_H^2=\mu_H^2+c_hT^2,\tilde{\mu}_S^2=\mu_S^2+c_sT^2$ and $T_n$ is the nucleation temperature and indicates the onset of the EWPT.
During the intermediate phase, the singlet acquires non-zero vev, which induces the trilinear coupling $\mu_{hhs}$ and enables the vertex correction to the decay $N_2\to H\ell_\alpha$. The scalar mass $m_S$ also evolves with the temperature enabling the decay $N_2\to N_1S$ and switching on the CP asymmetry contributions from the RHN decays.
The evolutions of the scalar mass ($m_S(T)$) and the trilinear coupling ($\mu_{hhs}$) for one benchmark parameter point are shown in Fig.~\ref{fig:thermal_evolution}. The solid blue line shows $m_S(T)$; the light green shaded region indicates where it falls below the decay threshold $m_{N_2}-m_{N_1}$, enabling the enhanced CP asymmetry. The solid orange line shows $\mu_{hhs}(T)$, which is nonzero only during the first-step EWPT, so the vertex correction is active only in this window. Importantly, the GW signal from the strong EWPT depends critically on the same thermal evolution governing leptogenesis, establishing the direct connection explored below.

\section{Leptogenesis and Gravitational Wave Signal}

As the Universe cools, thermal corrections to the scalar mass $m_S(T)$, induced by $\lambda_{SH}$, can make the decay $N_2\to N_1 S$ kinematically accessible around the EWPT. This opens new sources of CP violation via self-energy and vertex corrections (cf. Fig.~\ref{fig:N_decay}), enhancing the baryon asymmetry. The corresponding asymmetries are given by~\cite{LeDall:2014too}:
\begin{align}
    \varepsilon_2^v &\sim \frac{|a_{21}|}{8\pi}\frac{\mu_{hhs}}{m_{N_2}}\sqrt{\frac{m_{N_1}}{m_{N_2}}}(\mathcal{F}_{jLL}^v + \mathcal{F}_{jRL}^v),\\
    \varepsilon_2^s &\sim \frac{|a_{11}a_{21}|}{8\pi}\sqrt{\frac{m_{N_1}}{m_{N_2}}}(\mathcal{F}_{jlLL}^s+\mathcal{F}_{jlRL}^s+\mathcal{F}_{jlLR}^s+\mathcal{F}_{jlRR}^s),
\end{align}
where the loop functions $\mathcal{F}^{v,s}_{\cdot}$ depend on $m_{N_{1,2}}$ and $m_S$. The trilinear coupling $\mu_{hhs}$, sourced during the first step of a two-step EWPT (see Eq.~\ref{eq:thermal}), directly links $\varepsilon_2^v$ to the thermal history of EWPT. The total CP asymmetries entering the Boltzmann equations are:
\begin{align}
    \varepsilon_1 = \varepsilon_1^0, \quad \varepsilon_2=\varepsilon_2^0+\varepsilon_2^v+\varepsilon_2^s,
\end{align}
with $\varepsilon_i^0$ the CP asymmetry contribution from standard type-I seesaw determined by $h_{\alpha i}$ in Eq.~\ref{eq:LagHE}.

The evolution of the RHN abundances $N_{N_i}$ and $B-L$ number $N_{B-L}$ follows the standard Boltzmann equations with scalar-induced decays and scattering included~\cite{LeDall:2014too},
\begin{align}
    \frac{\mathrm{d}N_{N_2}}{\mathrm{d}z} =& -(D_2 + D_{21})\Delta_2 + D_{21}\Delta_1\nonumber \\
    &\quad -\Delta_{12}S_{N_1N_2\to HH}-\Delta_{22}S_{N_2N_2\to HH}\\
    \frac{\mathrm{d}N_{N_1}}{\mathrm{d}z}=&- (D_1+D_{21})\Delta_1 + D_{21}\Delta_2 \nonumber\\
    &\quad -\Delta_{12}S_{N_1N_2\to HH} - \Delta_{11}S_{N_1N_2\to HH}\\
    \frac{\mathrm{d}N_{B-L}}{\mathrm{d}z} =& -\sum_{i=1}^2\varepsilon_iD_i\Delta_i - WN_{B-L}
\end{align}
where $\Delta_i=\frac{N_{N_i}}{N_{N_i}^{\rm eq}}-1$ and $\Delta_{ij} = \frac{N_{N_i}N_{N_j}}{N_{N_i}^{\rm eq}N_{N_j}^{\rm eq}}-1$. The decay and washout terms are
\begin{gather}
  D_i (z)= K_i\,z \,\frac{\mathcal{K}_1(z_i)}{\mathcal{K}_2(z_i)} N_{N_i}^{\text{eq}}(z)\,,\\
  D_{21}(z) = K_{21} \,z\,\frac{\mathcal{K}_1(z_2)}{\mathcal{K}_2(z_2)} N_{N_2}^{\text{eq}}(z)\,,\\
  W(z) = \sum_i\frac{1}{4}K_iz_i^3\mathcal{K}_1(z_i)\,,
\end{gather}
where $z_i = m_{N_i}/T$, $z \equiv z_1$ and $\mathcal{K}_i(z_i)$ are the $i$-th modified Bessel function of the second kind. The decay parameters are determined by the corresponding decay width and the Hubble parameter:
\begin{equation}\label{eq:decay_parameter}
K_i \equiv \frac{\Gamma(N_i\to L H)}{H(T=m_{N_i})},\quad K_{21}  \equiv \frac{\Gamma(N_2\to N_1 S)}{H(T=m_{N_2})}.
\end{equation}
The baryon asymmetry freezes at the sphaleron decoupling temperature, identified with the EWPT nucleation temperature $T_n$, and is given by
\begin{align}
    \eta_B \simeq 0.013\times N_{B-L}(T_n).
\end{align}

\begin{figure}[t!]
    \centering
   \includegraphics[width=0.45\textwidth]{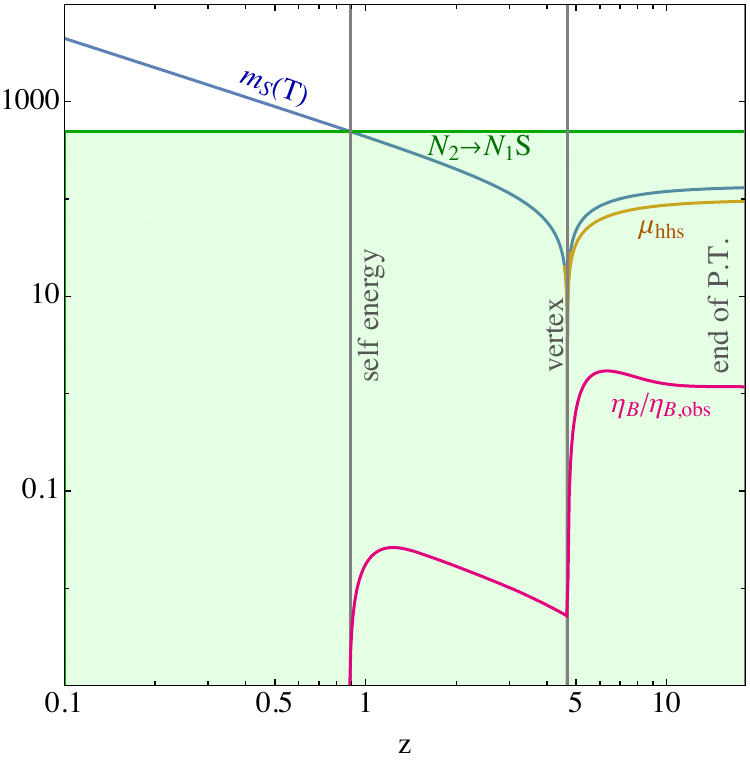}
    \caption{Evolution of the singlet mass $m_S$ (blue, GeV), trilinear coupling $\mu_{hhs}$ (orange, GeV), and baryon asymmetry $\eta_B/\eta_{B,\rm obs}$ (magenta) for the benchmark $\lambda_S \approx 0.382$, $\lambda_{SH} \approx 0.317$, $|a_{21}| = 0.01$, $|a_{11}| = 10^{-4}$, $m_{S}(0) = 100$~GeV, $m_{N_1} = 1000$~GeV, $m_{N_2} = 1500$~GeV. The light green region marks where $m_S(T) < m_{N_2} - m_{N_1}$, enabling the $N_2\to N_1 S$ decay.
    }
    \label{fig:thermal_evolution}
\end{figure}

The scalar sector impacts leptogenesis in two crucial ways: (i) temperature-dependent thermal mass $m_S(T)$ determines the threshold of $N_2\to N_1 S$ decays, controlling the onset of $\varepsilon_2^{v,s}$; (ii) the trilinear coupling $\mu_{hhs}(T)$, nonzero only during the first-step EWPT, also triggers the contribution of $\varepsilon_2^v$. These are illustrated in Fig.~\ref{fig:thermal_evolution}, where the magenta line shows the evolution of $\eta_B/\eta_{B,\rm obs}$.

In the absence of scalar effects, the CP asymmetry from type-I seesaw mechanism is $|\varepsilon_i^0|\lesssim 10^{-14}$ for our benchmark with light RHNs yields negligible baryon asymmetry. When scalar-induced processes are included, the evolution of $\eta_B$ clearly exhibits two distinct contributions from the scalar sector discussed above. In this benchmark, near $z\approx 0.90$, $N_2\to N_1S$ opens, triggering $\varepsilon_2^s\sim 10^{-8}$; later, at $z\approx 4.7$, the EWPT induces $\mu_{hhs}(T)$, activating $\varepsilon_2^v\sim 10^{-5}$. This synergy enables successful low-scale leptogenesis correlated with the dynamics of the EWPT.

Crucially, the period during which $\mu_{hhs}(T)$ remains non-zero is controlled by the duration of the EWPT, terminated at $T_n$. A lower $T_n$ extends this window and allows for greater accumulation of the asymmetry. Simultaneously, a lower $T_n$ indicates a stronger first-order EWPT, resulting in more intense GW signals.  This establishes a direct link between the BAU yield $\eta_B$ and the detectability of GWs.

To explore this interplay, we perform a scan in the scalar sector over $(\lambda_S,\lambda_{SH})$ while fixing $m_S(T=0)=100\,\rm GeV$. The BAU is calculated further fixing $m_{N_1}=1000$ GeV, $m_{N_2}=1500$ GeV, $|a_{21}|=0.01$, $|a_{11}|=10^{-4}$, $K_1=12$ and $K_2=20$. The conventional seesaw contribution $\varepsilon_i^0$ and the scattering terms are negligible in this setup. The robustness of the resulting correlation against variations of these
benchmark parameters is explicitly demonstrated in
the supplemental material.
The GW spectrum is computed using lattice-calibrated fitting formulas~\cite{Jinno:2016vai,Huber:2008hg,Kamionkowski:1993fg,Hindmarsh:2013xza,Hindmarsh:2015qta,Hindmarsh:2017gnf,Caprini:2015zlo,Caprini:2019egz,Ellis:2018mja,Guo:2020grp,Espinosa:2010hh,Caprini:2009yp,Binetruy:2012ze} with bubble wall velocity $v_w=0.6$. The detailed calculations of GW spectrum and the dependence on the choices of $v_w$ are also provided in the supplemental material.

\begin{figure}[t!]
    \centering
    \includegraphics[width=0.43\textwidth]{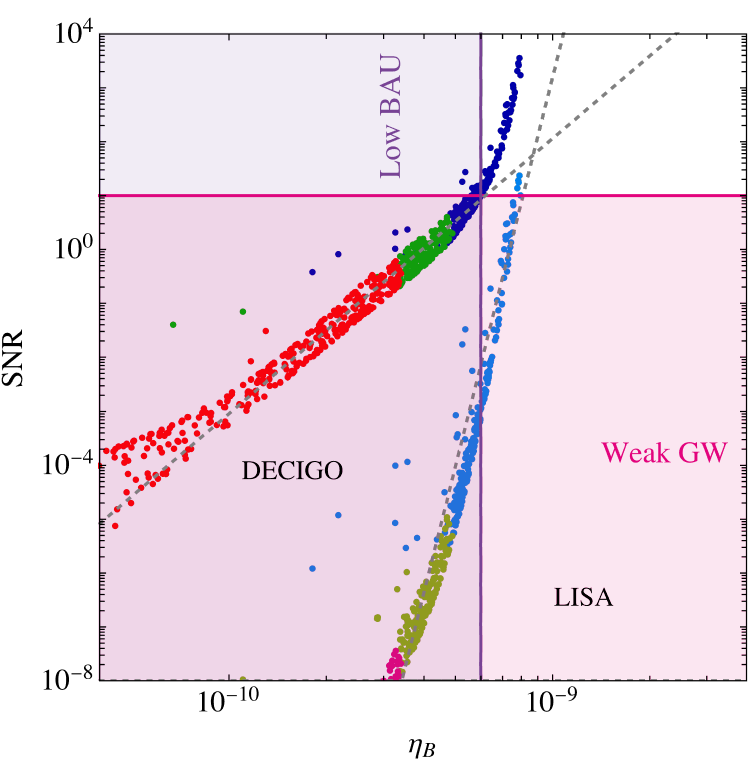}
    \caption{Correlation between baryon asymmetry $\eta_B$ and the SNR of the GW signal for both LISA (lighter points) and DECIGO (darker points). The red, green and blue points correspond to $T_c-T_n\in [0,5]$, $[5,10]$ and $[10,\infty]$ GeV, respectively.
    }
    \label{fig:BAU_GW}
\end{figure}

As shown in Fig.~\ref{fig:BAU_GW}, the baryon asymmetry and GW SNR are strongly correlated, both scaling with $\lambda_{S,SH}$, which governs $\mu_{hhs}$ and hence both the CP source and the strength of the EWPT. A small fraction of the points with strong enough EWPT ($T_c-T_n>10$ GeV) yield detectable signals for LISA~\cite{Caprini:2015zlo} with $\text{SNR}>10$ and simultaneously generate sufficient $\eta_B$. The superior sensitivity of U-DECIGO~\cite{Kawamura:2011zz,Kawamura:2006up} extends this coverage to more first-order EWPT points, as shown with darker markers in Fig.~\ref{fig:BAU_GW}. The linear trend in the $\eta_B$-SNR plane reflects the common scalar origin of the CP asymmetry and GW productions. This tight correlation between low-scale leptogenesis and GW observables presents a unique testable framework where future GW detectors not only probe the EWPT but could also verify the mechanism behind the baryon asymmetry of the Universe.
\section{Testing Leptogenesis from Observation of Gravitational Waves}
\label{sec:testing}

The correlation shown in Fig.~\ref{fig:BAU_GW} demonstrates that $\eta_B$ and the GW SNR move together, but does not by itself specify what a GW measurement would actually test. Future collider or displaced-vertex searches, together with neutrino oscillation data, could in principle pin down the RHN-sector parameters entering the decays $N_2\to \ell H$ and $N_2\to N_1 S$, namely $m_{N_{1,2}}$, $a_{21}$, $K_{1,2}$, and $m_S(T=0)$. However, the two scalar self-couplings $\lambda_S$ and $\lambda_{SH}$ that govern the EWPT dynamics are not directly fixed by these measurements; they require an independent probe of the scalar potential itself, e.g.\ via double/triple Higgs couplings, or via the EWPT dynamics they induce.

\begin{figure}[t!]
    \centering
   \includegraphics[width=0.45\textwidth]{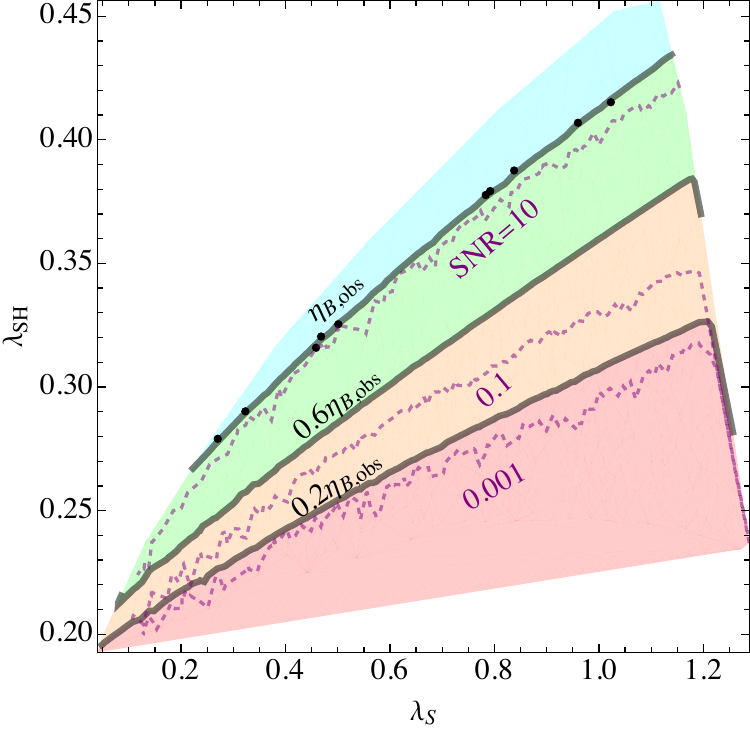}
    \caption{Dependence of the baryon asymmetry $\eta_B$ and the GW signal-to-noise ratio  on the scalar couplings $\lambda_S$ and $\lambda_{SH}$, for the fixed RHN-sector parameters of Fig.~\ref{fig:thermal_evolution}.}
    \label{fig:lam_BAU_SNR}
\end{figure}

This is precisely where the shared origin identified above becomes a quantitative test rather than a qualitative analogy. For fixed RHN-sector parameters, the same couplings $(\lambda_S,\lambda_{SH})$ that set the strength and duration of the EWPT (and hence the GW amplitude, frequency, and SNR) also set the size and duration of the extra CP-violating sources $\varepsilon_2^{v,s}$ (and hence $\eta_B$), through the common thermal history of Eq.~\eqref{eq:thermal}. This is illustrated in Fig.~\ref{fig:lam_BAU_SNR}: requiring the observed asymmetry $\eta_{B,\rm obs}$ selects a specific band in the $(\lambda_S,\lambda_{SH})$ plane, while the GW SNR varies systematically over the same plane. Scanning over $(\lambda_S,\lambda_{SH})$ at fixed RHN parameters, as done in Fig.~\ref{fig:BAU_GW}, therefore traces out a specific correlation between the predicted $\eta_B$ and the GW SNR, which we have checked to be robust against variations of the RHN-sector parameters in the Supplemental Material.

This correlation converts a GW measurement into a direct, two-sided statement about leptogenesis in our framework, for a given, independently measured RHN sector: (i) if a future detection yields a GW SNR and spectral shape whose inferred $(\lambda_S,\lambda_{SH})$ fall inside the band reproducing $\eta_{B,\rm obs}$, this constitutes highly non-trivial supporting evidence for the mechanism, since nothing in the theory otherwise forces the phase-transition strength and the baryon asymmetry to coincide; (ii) conversely, a null search at LISA/U-DECIGO in the relevant frequency band would exclude the region of $(\lambda_S,\lambda_{SH})$ required to reproduce $\eta_{B,\rm obs}$ within this mechanism, thereby excluding scalar-assisted leptogenesis as the origin of the observed asymmetry for that RHN spectrum. In either case, the GW measurement---positive or null---carries genuine, falsifiable information about the viability of leptogenesis in our framework, unlike in generic scalar-extended models where the phase-transition sector and the baryogenesis sector are mutually independent and a GW measurement carries no information about the BAU mechanism. This is the sense in which we regard GW observations as a probe, rather than merely an observational avenue, for leptogenesis.

\section{Conclusion}

We have demonstrated that scalar-assisted leptogenesis provides a compelling solution to the challenges of conventional scenarios. By extending the scalar sector with particles that couple to both RHNs and the SM Higgs, we achieve successful low-scale leptogenesis with light RHNs and without fine-tuned mass degeneracy, while simultaneously inducing a strong first-order EWPT that generates observable GW signals. Importantly, our work establishes a direct multi-messenger connection: the baryon asymmetry and gravitational-wave signals originate from the same scalar-RHN dynamics, providing a novel testable pathway to probe leptogenesis through future GW observations. This represents a fundamental shift from conventional approaches where these observables are typically independent.

A key result is the intrinsic correlation between the baryon asymmetry $\eta_B$ and the strength of the GW signal. Achieving detectable GWs requires a strong first-order EWPT with a low nucleation temperature $T_n$, driven by large scalar couplings. The same couplings control the temperature evolution of the scalar mass and trilinear interactions, which govern the kinematics and size of the CP asymmetry during leptogenesis. Consequently, the parameter space yielding sufficient BAU is tightly aligned with that predicting observable GWs, establishing a direct experimental link between leptogenesis and gravitational wave astronomy. We stress that neither the shape of the GW spectrum in isolation, nor the discovery of the singlet scalar or the RHNs at colliders, would by itself establish a connection to leptogenesis. What makes the connection genuine and falsifiable is the bidirectional logic detailed above: for a given, independently measurable RHN sector, the same couplings required to reproduce the observed baryon asymmetry also predict a specific GW signal, so that a future detection or non-detection directly supports or excludes scalar-assisted leptogenesis as the origin of the baryon asymmetry. This constitutes a qualitatively new type of multi-messenger connection between particle cosmology and gravitational-wave astronomy, in which the same underlying dynamics is simultaneously responsible for two independently measurable cosmological observables---a connection that is particularly timely given the imminent operation of LISA, Taiji, TianQin, and DECIGO-class missions.

Our framework offers promising avenues for future exploration. In particular, scenarios with supercooled EWPTs, where prolonged transitions can further amplify both the GW signal and the baryon asymmetry, deserve detailed investigation. Studies of RHN Yukawa coupling textures and their impact on the BAU-GW correlation will be pursued in future work. Ultimately, our results demonstrate that future GW detectors such as LISA and DECIGO
will provide crucial experimental verification of the leptogenesis mechanism,
establishing a novel probe of the baryon asymmetry origin within the coming decade.

\acknowledgments
We would like to thank Ke-Pan Xie for helpful discussion. W.L. is supported by National Natural Science foundation of China (Grant No. 12205153). Y.W. is supported by National Natural Science foundation of China under (Grant No. 12305112). The authors gratefully acknowledge the valuable discussions and insights provided by the members of the China Collaboration of Precision Testing and New Physics (CPTNP).
 \appendix
\section{Computation of the GW spectrum}

The stochastic gravitational wave (GW) background generated during the electroweak phase transition (EWPT) originates from three primary sources: bubble wall collisions, sound waves in the plasma, and magnetohydrodynamic (MHD) turbulence. The resulting spectrum depends principally on the macroscopic thermodynamic parameters of the transition: the phase transition strength parameter $\alpha$ (representing the ratio of latent heat to radiation energy density), the inverse duration of the transition relative to the Hubble scale $\beta/H_*$, and the transition temperature $T_*$. Regarding the bubble wall velocity $v_w$, we treat it as a phenomenological parameter, adopting benchmark values that cover the subsonic ($v_w < c_s$), sonic ($v_w \approx c_s$), and supersonic ($v_w > c_s$) expansion regimes, as well as Chapman-Jouguet velocity as function of $\alpha$~\cite{Steinhardt:1981ct}:
\begin{align}
    v_w = \frac{\sqrt{1/3}+\sqrt{\alpha^2+2\alpha/3}}{1+\alpha}.
\end{align}

{\noindent\bf Bubble Collision Contribution:}
The gravitational wave spectrum arising from the collision of bubble walls is modeled using the envelope approximation. The spectrum is characterized by the peak amplitude and frequency scaling as~\cite{Jinno:2016vai}:
\begin{align}
\Omega_{\rm coll}h^2 &= 1.67\times 10^{-5}\left( \frac{g_{*}}{100} \right)^{-1/3}\left( \frac{\kappa_{\rm coll}\alpha}{1+\alpha} \right)^{2}\left( \frac{\beta}{H_{*}} \right)^{-2} \nonumber \\
&\times \Delta(v_{w})S_{\rm coll}(f,v_{w}),\\
f_{\rm coll} &= 1.65\times 10^{-5}\mathrm{Hz}\left( \frac{g_{*}}{100} \right)^{1/6}\left( \frac{T_{*}}{100\,\mathrm{GeV}} \right)\left( \frac{\beta}{H_{*}} \right)\left( \frac{f_{*}}{\beta} \right).
\end{align}
The spectral shape function $S_{\rm coll}(f)$ and the velocity-dependent normalization $\Delta(v_w)$ are given by:
\begin{align}
&S_{\rm coll}(f) \\ \nonumber
&= \left[ c_{l}\left( \frac{f}{f_{\rm coll}} \right)^{-3} + (1-c_{l}-c_{h})\left( \frac{f}{f_{\rm coll}} \right)^{-1}  
 + c_{h}\left( \frac{f}{f_{\rm coll}} \right) \right]^{-1} \\ 
&\Delta(v_{w}) = \frac{0.48v_{w}^{3}}{1+5.3v_{w}^{2}+5.0v_{w}^{4}}
\end{align}

where the spectral fit parameters are $c_l \approx 0.064$ and $c_h \approx 0.48$. The peak frequency is determined by the factor:
\begin{equation}
\frac{f_{*}}{\beta} = \frac{0.35}{1+0.069v_{w}+0.69v_{w}^{4}}.
\end{equation}
Note that other fitting is also available in the literature~\cite{Huber:2008hg}. The amplitude is also governed by the efficiency factor $\kappa_{\rm coll}$, representing the fraction of latent heat transformed into the kinetic energy of the bubble shell. An analytic fit for this efficiency is given as~\cite{Kamionkowski:1993fg}:
\begin{equation}
\kappa_{\rm coll} = \frac{1}{1+0.715\alpha}\left( 0.715\alpha + \frac{4}{27}\sqrt{ \frac{3\alpha}{2} } \right).
\end{equation}
It is crucial to note that the efficiency $\kappa_{\rm coll}$ depends heavily on the expansion regime. In the runaway regime, the bubble walls accelerate continuously, allowing $\kappa_{\rm coll} \to 1$. However, for the thermal phase transitions considered in this work, the interaction between the bubble wall and the plasma typically generates sufficient friction to force the wall into a terminal velocity (the non-runaway regime)~\cite{Bodeker:2009qy,Bodeker:2017cim}. In this state, the vast majority of the vacuum energy is transferred to the plasma fluid motion (sound waves) rather than accelerating the wall itself. Consequently, $\kappa_{\rm coll}$ becomes negligible ($\kappa_{\rm coll} \ll \kappa_{\rm sw}$), making the bubble collision contribution subdominant to the sound wave and turbulence signals~\cite{Hindmarsh:2013xza,Caprini:2015zlo,Caprini:2019egz}. Therefore, we neglect $\Omega_{\rm coll}h^2$ in our calculation and focus on the dominant sound wave and turbulence sources discussed below.

{\noindent\bf Sound Wave Contribution:}
The dominant source of gravitational waves in non-runaway phase transitions is typically the stochastic background produced by sound waves (bulk fluid motion) in the plasma. Based on numerical simulations and the sound shell model~\cite{Hindmarsh:2013xza,Hindmarsh:2015qta,Hindmarsh:2017gnf}, the resulting GW power spectrum is given by:
\begin{align}
\Omega_{\rm sw}h^{2} &= 2.59\times 10^{-6} \left( \frac{g_{*}}{100} \right)^{-1/3}\left( \frac{\kappa_{\rm sw}\alpha}{1+\alpha} \right)^{2} \left( \frac{\beta}{H_{*}} \right)^{-1} \nonumber \\
&\times \Delta(v_w) \Upsilon(\tau_{\rm sw}) S_{\rm sw}(f), \\
f_{\rm sw} &= 8.9\times 10^{-6},\mathrm{Hz} \left( \frac{g_{*}}{100} \right)^{1/6}\left( \frac{T_{*}}{100,\mathrm{GeV}} \right) \frac{1}{\Delta(v_w)} \left( \frac{\beta}{H_{*}} \right),
\end{align}
where we have defined the velocity factor $\Delta(v_w) \equiv \max(v_{w},c_{s})/c$ to account for the relevant length scale of the source, and $c_{s}\approx 1/\sqrt{3}$ is the sound speed in the plasma. The spectral shape function $S_{\rm sw}(f)$ follows a broken power law:
\begin{align}
    S_{\rm sw}(f) = \left( \frac{f}{f_{\rm sw}} \right)^{3}\left( \frac{7}{4+3\left( f/f_{\rm sw} \right)^{2}} \right)^{7/2}.
\end{align}
A critical correction to the standard sound wave spectrum is the suppression factor $\Upsilon(\tau_{\rm sw})$, which accounts for the finite lifetime of the acoustic source. Sound waves do not persist indefinitely; they eventually develop non-linearities and form shocks, leading to the cessation of the acoustic phase and the onset of turbulence. The suppression is modeled as~\cite{Ellis:2018mja,Guo:2020grp}:
\begin{align}
    \Upsilon(\tau_{\rm sw}) = 1 - \frac{1}{\sqrt{1+2H_*\tau_{\rm sw}}},
\end{align}
where $\tau_{\rm sw}$ is the duration of the acoustic phase (the shock formation timescale). This timescale is estimated by the ratio of the mean bubble separation $R_*$ to the root-mean-square fluid velocity $\bar{U}_f$:
\begin{align}
    \tau_{\rm sw} \approx \frac{R_*}{\bar{U}_f}, \quad \text{with} \quad R_* = (8\pi)^{1/3}\frac{v_w}{\beta}, \quad \bar{U}_f = \sqrt{\frac{3}{4}\frac{\kappa_{\rm sw}\alpha}{1+\alpha}}.
\end{align}
This factor ensures that if the shock formation time is shorter than the Hubble time ($H_*\tau_{\rm sw} < 1$), the total GW power is reduced accordingly.

The amplitude of the signal also depends on the efficiency factor $\kappa_{\rm sw}$, which represents the fraction of vacuum energy converted into bulk fluid motion. We utilize the kinetic energy fraction fits provided by Espinosa et al.~\cite{Espinosa:2010hh}, covering the subsonic (deflagration), sonic (Jouguet), and supersonic (detonation) regimes:
\begin{align}
    \kappa_{\rm sw} \simeq \begin{cases}
        &\frac{c_{s}^{11/5}\kappa_{A}\kappa_{B}}{(c_{s}^{11/5}-v_{w}^{11/5})\kappa_{B}+v_{w}c_{s}^{6/5}\kappa_{A}} \quad v_{w} \lesssim c_{s}  (\text{Subsonic}), \\
        &\kappa_{B}+(v_{w}-c_{s})\delta \kappa 
        + \frac{(v_{w}-c_{s})^{3}}{(c_{J}-c_{s})^{3}}\left[\kappa_{C}-\kappa_{B}-(c_{J}-c_{s})\delta \kappa\right] \\ &c_{s} < v_{w} < c_{J}  (\text{Sonic}), \\
        &\frac{(c_{J}-1)^{3}c_{J}^{5/2}v_{w}^{-5/2}\kappa_{C}\kappa_{D}}{((c_{J}-1)^{3}-(v_{w}-1)^{3})c_{J}^{5/2}\kappa_{C}+(v_{w}-1)^{3}\kappa_{D}} \\ &v_{w} \gtrsim c_{J} \quad (\text{Supersonic}),
    \end{cases}
\end{align}
where the Chapman-Jouguet velocity is $c_{J} = (\sqrt{ \frac{2}{3}\alpha+\alpha^{2} }+\sqrt{1/3})/(1+\alpha)$ and the crossover correction is $\delta\kappa \approx -0.9\log(\sqrt{\alpha}/(1+\sqrt{\alpha}))$. The boundary coefficients are given by:
\begin{align}
    \kappa_{A} &\simeq v_{w}^{6/5} \frac{6.9\alpha}{1.36-0.037\sqrt{ \alpha }+\alpha}, \\
    \kappa_{B} &\simeq \frac{\alpha^{2/5}}{0.017 + (0.997+\alpha)^{2/5}}, \\
    \kappa_{C} &\simeq \frac{\sqrt{ \alpha }}{0.135+\sqrt{ 0.98+\alpha }}, \\
    \kappa_{D} &\simeq \frac{\alpha}{0.73 + 0.083\sqrt{ \alpha }+\alpha}.
\end{align}

{\noindent\bf MHD Turbulence Contribution:}
As the sound wave contribution described above proceeds, non-linearities in the fluid motion eventually become significant. The sound waves decay, generating shocks and subsequently leading to Magnetohydrodynamic (MHD) turbulence in the plasma. This turbulence acts as a secondary source of gravitational waves, which can persist well after the acoustic source has effectively shut off. The resulting stochastic GW background is modeled analytically as~\cite{Caprini:2009yp,Binetruy:2012ze}:
\begin{align}
\Omega_{\rm turb}h^{2} &= 3.35\times 10^{-4} \left( \frac{g_{*}}{100} \right)^{-1/3}\left( \frac{\beta}{H_{*}} \right)^{-1} \left( \frac{\kappa_{\rm turb}\alpha}{1+\alpha} \right)^{3/2} \nonumber \\
&\times \Delta(v_w) S_{\rm turb}(f), \\
f_{\rm turb} &= 2.7\times 10^{-5},\mathrm{Hz}\left( \frac{g_{*}}{100} \right)^{1/6}\left( \frac{T_{*}}{100,\mathrm{GeV}} \right) \\ \nonumber &\times
\frac{1}{\Delta(v_w)} \left( \frac{\beta}{H_{*}} \right).
\end{align}
Consistent with the sound wave treatment, the characteristic velocity scale is set by $\Delta(v_w) = \max(v_{w},c_{s})/c$. The spectral shape $S_{\rm turb}(f)$ is determined by the Kolmogorov theory of turbulence, modified by a cutoff at the Hubble scale to respect causality. It is given by:
\begin{align}
S_{\rm turb}(f) = \frac{\left( f/f_{\rm turb} \right)^{3}}{\left[ 1+\left( f/f_{\rm turb} \right) \right]^{11/3} \left( 1+ 8\pi f/h_{*} \right)},
\end{align}
where the peak frequency $f_{\rm turb}$ corresponds to the inverse of the mean bubble separation scale ($R_*$). The term $h_*$ represents the redshifted Hubble rate at the time of generation, which sets the horizon scale:
\begin{equation}
h_* = 1.65 \times 10^{-5}\,\mathrm{Hz}\left( \frac{T_{*}}{100\,\mathrm{GeV}} \right)\left( \frac{g_{*}}{100} \right)^{1/6}.
\end{equation}
At high frequencies ($f \gg f_{\rm turb}$), the spectrum follows the characteristic Kolmogorov slope ($\Omega \sim f^{-5/3}$), while the term $(1+ 8\pi f/h_{*})$ ensures the correct asymptotic behavior at scales approaching the horizon.

The amplitude of this signal is governed by the turbulence efficiency factor $\kappa_{\rm turb}$, which represents the fraction of total vacuum energy converted into vortical fluid motion. Recent numerical simulations of the EWPT suggest that the conversion from bulk fluid motion (sound waves) to turbulence is not fully efficient. It is generally estimated that a fraction $\epsilon \sim 5\% - 10\%$ of the bulk kinetic energy is transferred to the turbulent cascade~\cite{Hindmarsh:2015qta,Caprini:2019egz}. Consequently, we adopt the standard approximation for our analysis:
\begin{equation}
    \kappa_{\rm turb} \approx \epsilon\kappa_{\rm sw} \approx 0.1 \kappa_{\rm sw}.
\end{equation}
While often subdominant to the sound wave signal in peak amplitude, the turbulence contribution can dominate the high-frequency tail of the spectrum due to its slower spectral decay compared to sound waves.

\begin{figure*}[t!]
\centering
\includegraphics[width=0.32\textwidth]{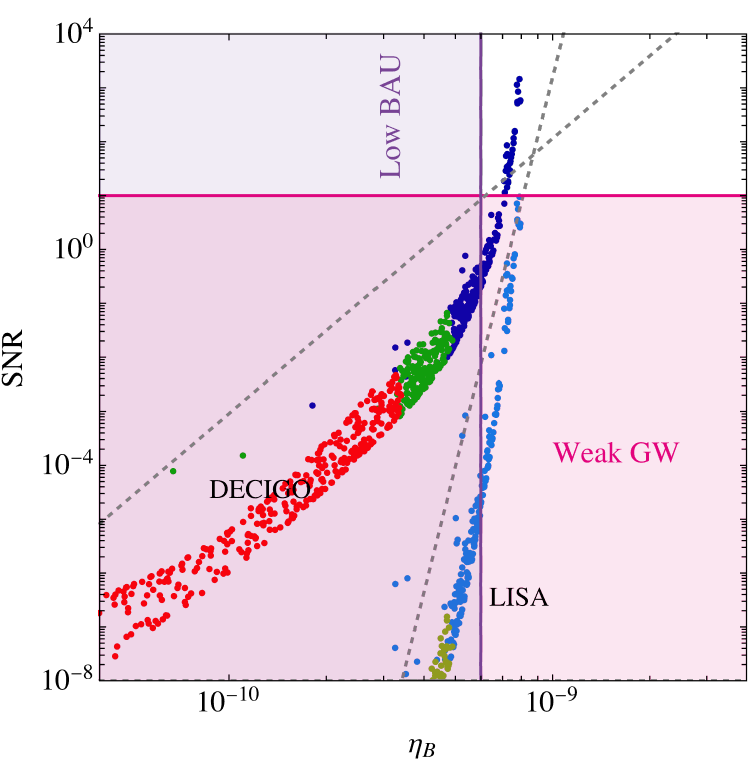}
\includegraphics[width=0.32\textwidth]{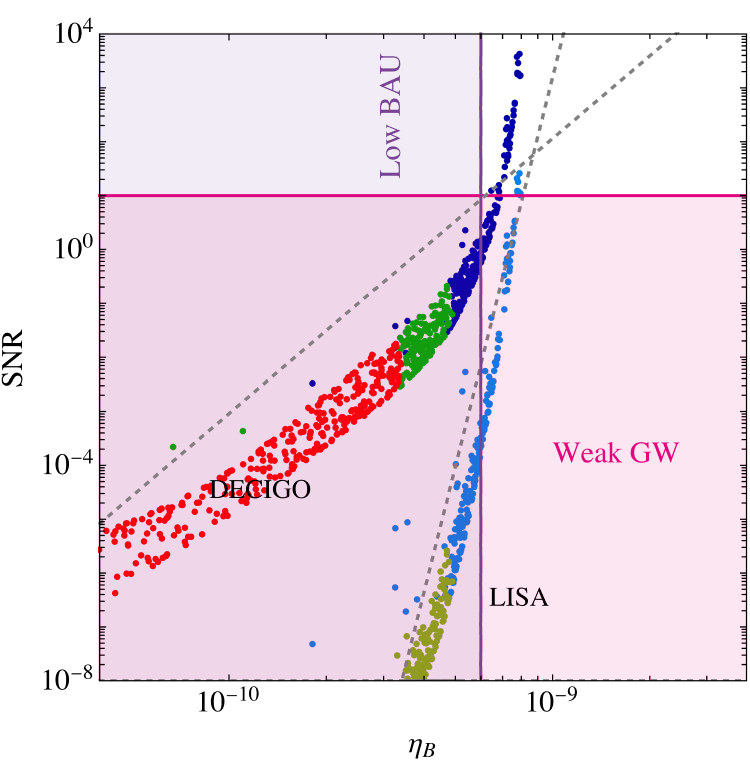}
\includegraphics[width=0.32\textwidth]{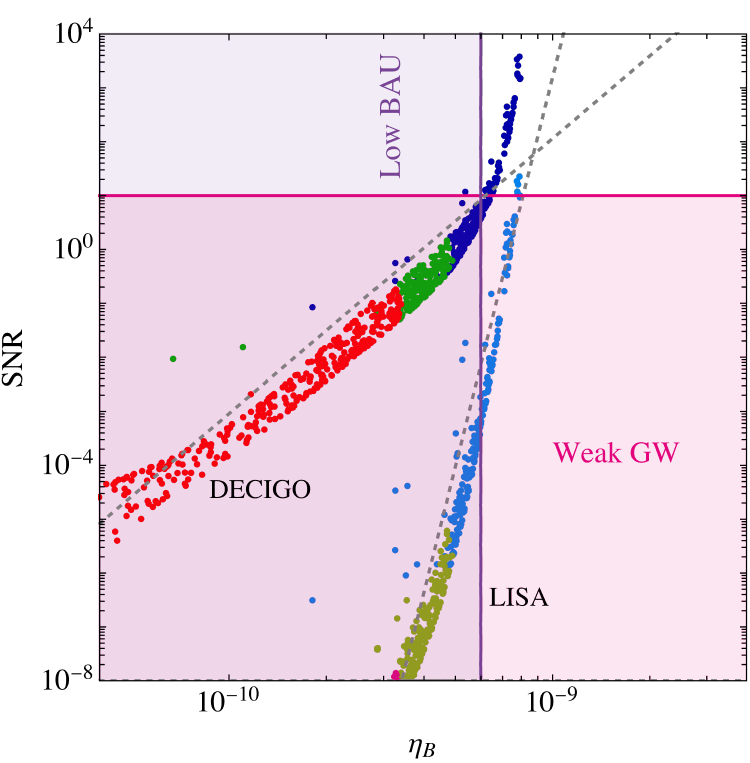}
\caption{
Same as Fig.~3 in the main text, but with variate calculations of GW:
(a) $v_w = 0.3$,
(b) $v_w = 0.9$,
(c) Chapman-Jouguet velocity,
}
\label{fig:vw}
\end{figure*}

In the main text, we presented the gravitational wave (GW) spectrum for a benchmark bubble wall velocity of $v_w=0.6$. In \autoref{fig:vw}, we extend this analysis to include a range of wall velocities, covering the subsonic, sonic, and supersonic regimes. As expected from the hydrodynamic relations discussed above, variations in $v_w$ affect the peak amplitude of the GW signal, with certain velocities leading to a suppression of the spectral strength (shifting the points downwards). However, the conclusion of our work remains robust against these variations. The strong correlation between the GW signal strength and the baryon asymmetry (BAU) generated via leptogenesis persists across different velocity benchmarks, demonstrating that this feature is a fundamental prediction of the model rather than an artifact of a specific parameter choice.

\section{Robustness of the $\eta_b$-GW SNR Correlation Under Individual Parameter Variations}

In this section, we demonstrate that the correlation between the baryon asymmetry $\eta_b$ and the gravitational wave signal-to-noise ratio (SNR) is robust under changes in individual model parameters.

\begin{figure*}[t!]
\centering
\includegraphics[width=0.24\textwidth]{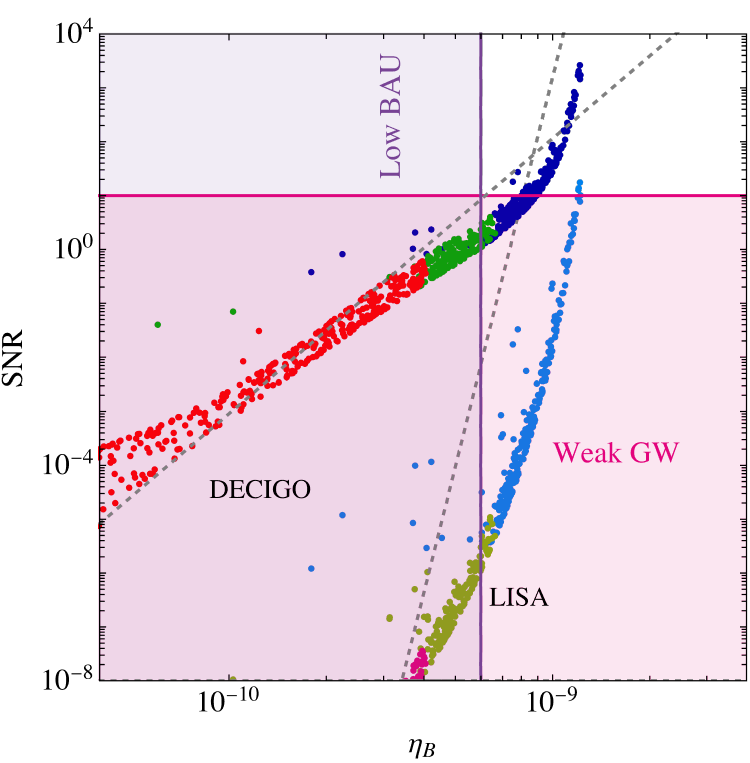}
\includegraphics[width=0.24\textwidth]{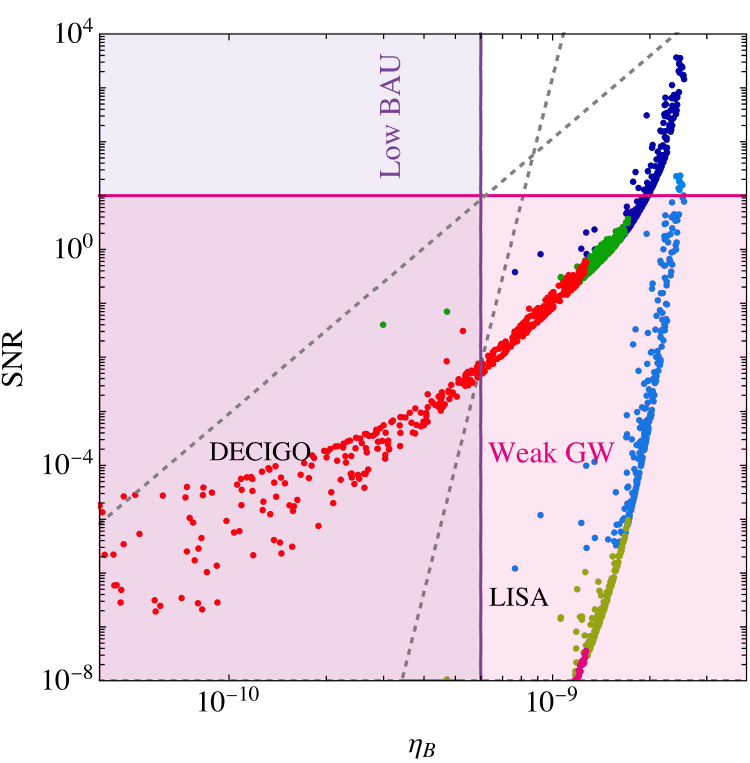}
\includegraphics[width=0.24\textwidth]{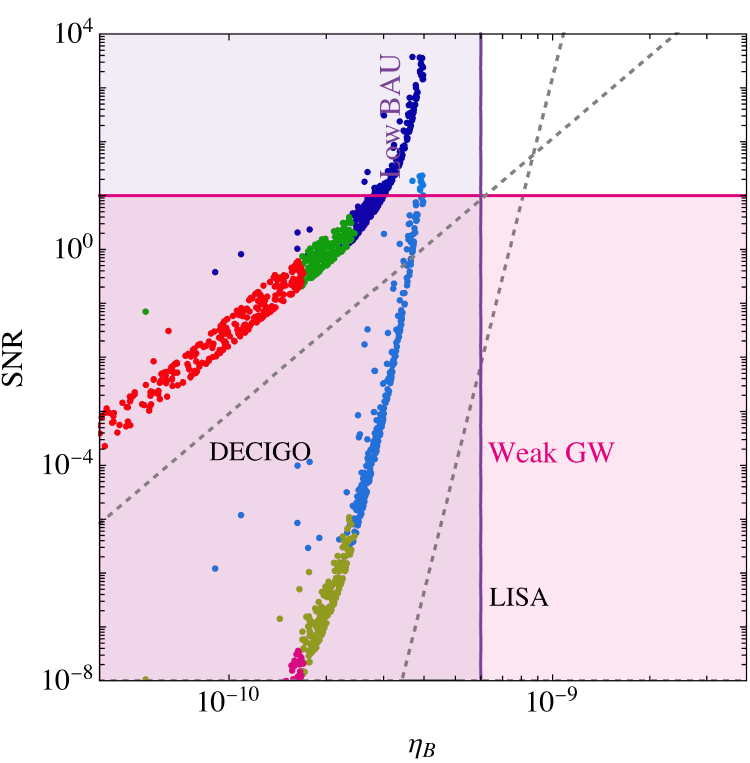}
\includegraphics[width=0.24\textwidth]{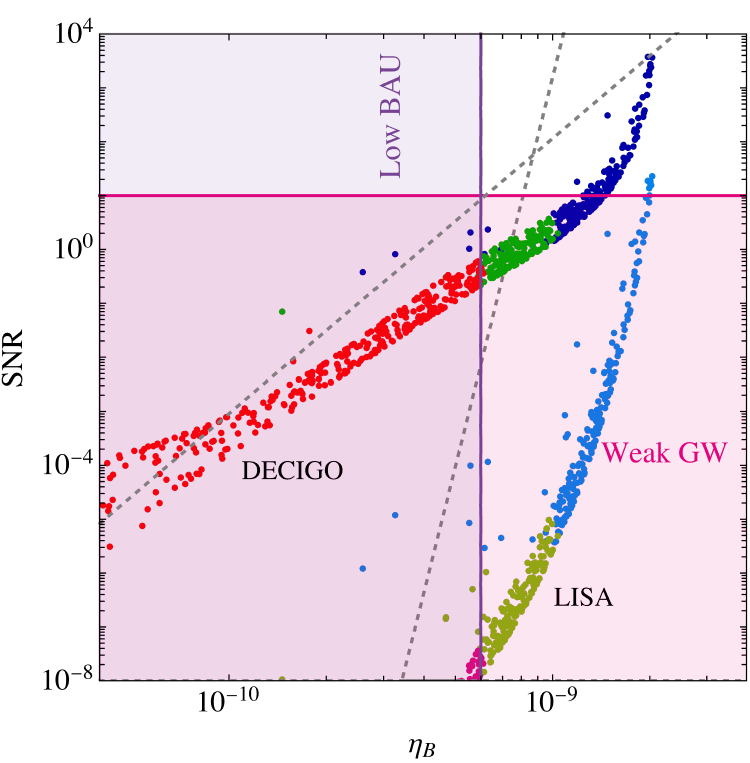}
\caption{
Same as Fig.~3 in the main text, but with individual parameter variations:
(a) $m_{N_1} = 1200~\mathrm{GeV}$,
(b) $m_{N_2} = 1200~\mathrm{GeV}$,
(c) $\alpha_{12} = 0.005$,
and (d) $(K_1, K_2) = (5, 10)$.
In each panel, only one parameter is varied while all others are fixed to the benchmark values.
The dashed lines roughly trace the original distribution.
}
\label{fig:vary}
\end{figure*}

In Fig.~\ref{fig:vary}, we vary one parameter at a time while keeping all others fixed at the benchmark values used in the main text:
\begin{itemize}
    \item $m_{N_1}$: varied from 1000 GeV to 1200 GeV in Fig.~\ref{fig:vary} (a)
    \item $m_{N_2}$: varied from 1500 GeV to 1200 GeV in Fig.~\ref{fig:vary} (b)
    \item $\alpha_{12}$: varied from $10^{-2}$ to 0.005 in Fig.~\ref{fig:vary} (c)
    \item $K_1$ and $K_2$: varied from (12, 20) to (5, 10) in Fig.~\ref{fig:vary} (d).
\end{itemize}

As shown in the figures, in all cases the qualitative trend observed in the benchmark scenario is maintained, although the slope and magnitude vary with the choice of parameters. In particular, increasing $m_{N_1}$ generally leads to larger baryon asymmetry $\eta_B$ and a reduced slope, since the $CP$ asymmetry increases while
the mass splitting $m_{N_2}-m_{N_1}$ becomes smaller and the decay channel $N_2 \rightarrow N_1 S$ opens at later times. Decreasing $m_{N_2}$ results in the similar behavior.
Decreasing $\alpha_{12}$ reduces the vertex correction contribution to the $CP$ asymmetry and subsequently the final BAU. Finally, smaller values of $K_1$ and $K_2$ weaken washout effects, leading to a enhancement of the baryon asymmetry across all parameter points.

Overall, these results show that the approximately linear dependence between the baryon asymmetry and the gravitational wave signal persists over a wide range of parameters, demonstrating that the observed correlation is a robust feature of the mechanism rather than an artifact of a single benchmark choice.

\bibliographystyle{bibstyle}
\bibliography{submit}
\end{document}